\documentclass[
 reprint,
 amsmath,amssymb,
aps,
]{revtex4-1}

\usepackage{caption}
\usepackage{subcaption}
\usepackage{amsmath}
\usepackage{graphicx}
\usepackage{dcolumn}
\usepackage{bm}
\usepackage{float}
\usepackage{color}
\usepackage{natbib}

\usepackage{multirow}
\usepackage{placeins }

\usepackage{hyperref}
\usepackage[mathlines]{lineno}

\begin{document}


\title{Adiabatic Floquet-Wave Expansion for the Analysis of Leaky-Wave Holograms Generating Polarized Vortex Beams}

\author{Amrollah Amini}
 \email{amini\_am@elec.iust.ac.ir}

\author{Homayoon Oraizi}%
 \email{h\_oraizi@iust.ac.ir}
\affiliation{%
 School of Electrical Engineering, Iran University of Science and Technology, 
 \\1684613114, Tehran, Iran
}%





\begin{abstract}
This paper presents the combination of aperture field estimation (AFE) technique and adiabatic Floquet-wave (AFW) expansion method for the analysis of leaky-wave holograms capable of generating orbital angular momentum (OAM) vortex waves. In these formulations the propagation and leakage constants are theoretically estimated and as a result the aperture field and far-zone field are calculated.
This theoretical approach significantly reduces the computational complexity in such a way that the analysis of holograms with large dimensions is possible with low memory requirements. 
Isotropic and anisotropic unit cells are used to realize the holograms, which anisotropic structures show better performance in the control of wave polarization. The holograms are designed to generate vortex beams with topological charge of m=2 and circular polarization states in the microwave regime. To evaluate the accuracy of the proposed method, the results of theoretical model and full-wave simulations are compared, with good agreement.
\end{abstract}

\pacs{Valid PACS appear here}
\maketitle


\section{\label{sec:introduction}Introduction}
Increasing the channel capacity of wireless systems is an essential demand in the implementation of new generations of telecommunication networks. An attractive way to meet this need without increasing the frequency bandwidth is to utilize photon orbital angular momentum (OAM) in the microwave regime \cite{thide2007}. Electromagnetic waves with helical spatial phase profiles can support different orthogonal OAM modes. Theoretically, the number of such modes is infinite. Consequently, combining these new multiplexing dimensions with other existing schemes such as time, frequency and polarization multiplexing can significantly increase the channel capacity  \cite{chen2020}. In recent years, various methods have been proposed to generate the radio-frequency waves carrying OAM modes. Utilizing phased array antennas with circular arrangements \cite{gao2014}, spiral phase plates  \cite{hui2015}, all-dielectric transformers \cite{yi2019, yi2_2019}, transmit-arrays \cite{jiang2018} and reflect-arrays \cite{karimipour2019} are among the most common methods for synthesizing OAM vortex waves. 
In phased array antennas the complexity of the feed network and the destructive effects of mutual coupling among adjacent elements are important limiting factors. Note that, eliminating the mutual couplings among elements in phased arrays drastically increases the complexity of realization process. 
To avoid these issues, all-dielectric transformers are proposed in the literature. In addition, reducing the divergence effects of vortex waves can be achieved by these transformers \cite{yi2_2019}. 
However, all-dielectric transformers and spiral phase plates are relatively bulky, which are unsuitable for use in low profile integrated systems. Recently, the advent of metasurfaces \cite{yu2011}-\cite{yu2014} has paved the way for the realization of integrated optical devices. These structures can manipulate electromagnetic wave front to obtain the desired properties by introducing abrupt phase shifts on the incident wave. Reflect/transmit arrays can be realized by metasurfaces. 
The use of metasurface-based reflect/transmit arrays has several advantages, including high gain, low divergence angles, low manufacturing cost and simple fabrication processes. However, in reflect/transmit arrays the feeding systems need to be mounted outside the structure, which makes them bulky. This issue contradicts the integrability of metasurfaces.
An alternative solution to achieve the advantages of reflect/transmit arrays without the need for bulky feed systems is utilizing the leaky-wave meta-holograms, in such a way that  the feed can be integrated with the metasurface plane. The application of holographic technique in  antenna engineering was first proposed by Checcacci \cite{checcacci1970} and a practical example of leaky-wave hologram was presented in \cite{fong2010}. 
In the holographic technique, the information obtained from the interference of the surface wave (as the reference wave) and the desired space wave (as the object wave) is utilized for the synthesis of metasurface. In \cite{oraizi2020} the isotropic hologram is used to generate radiations capable of carrying OAM modes. In \cite{amini_2020}-\cite{bodehou_OAM_2019} anisotropic holograms are designed as polarized vortex beam generators.
A serious obstacle in designing and optimizing leaky-wave holographic antennas is their large size. Generally, the dimensions of such structures to achieve a high gain radiation pattern should be selected about $10\lambda$ to $20\lambda$ ($\lambda$ as the free-space wavelength). On the other hand, due to the small size of unit cells in terms of wavelength ($\approx \lambda/6$), the number of meshes in full-wave simulations may increase significantly. This issue greatly increases the computational complexity and CPU time consumption.
Various theoretical models have been reported for the analysis and synthesis of leaky-wave holograms.
The aperture field estimation (AFE) method is proposed as an accurate synthesis method for realizing leaky-wave holograms with circular  \cite{casaletti_2017} and vertical polarizations \cite{amini_2020}.
In \cite{ovejero2015, bodehou2019}, the Method of Moments (MoM) framework is developed for the analysis of anisotropic holographic antennas, which is accurate and fast for treating large size structures. This method has been used to explain the radiation mechanism of holograms with circularly polarized shaped beams  \cite{minatti_2015}, multiple beams \cite{ovejero2017, bodehou2020} and multi band radiations \cite{faenzi2019}.
Another analytical method that can be considered is the adiabatic Floquet-wave expansion (AFW) method, which is relatively simpler than MoM in terms of mathematical complexity. This method is proposed for the first time by Minatti et al. \cite{minatti_2016, minatti_2_2016} for the shaping of far-zone patterns with different polarizations.

In this paper, the combination of aperture field estimation (AFE) technique (as  a synthesis method) and  adiabatic Floquet-wave (AFW) expansion method (as an analysis approach) is utilized to get deep insight into the radiation mechanism of OAM wave radiators enabled by leaky-wave holograms. Using AFE technique, the aperture field (consisting of both phase and amplitude information)  and then impedance distribution for obtaining the polarized vortex beam have been estimated. Furthermore, AFW  method  determines the leakage properties of the surface wave. 
This method is much faster than the full-wave method, which makes it suitable for synthesizing and optimizing complex vortex beams.
\section{Adiabatic floquet-wave expansion}
\begin{figure}
\begin{subfigure}[b]{0.5\textwidth}
	\includegraphics[width = \textwidth]{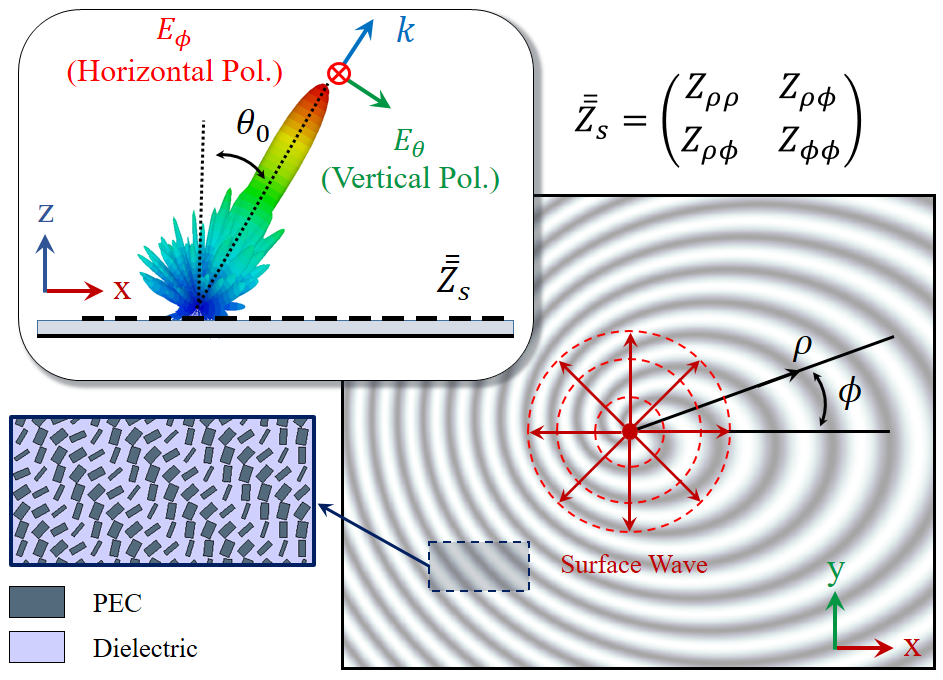}
	\caption{}
	\label{fig:Fig1a}
\end{subfigure}
\begin{subfigure}[b]{0.4\textwidth}
	\includegraphics[width = \textwidth]{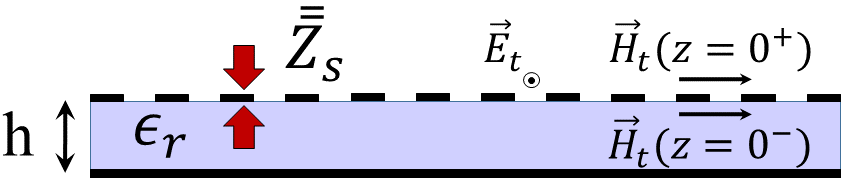}
	\caption{}
	\label{fig:Fig1b}
\end{subfigure}
\caption{(a) Conceptual schematic of anisotropic hologram consisting of monopole located at the origin. (b) Definition of transparent anisotropic surface impedance in presence of dielectric host medium.}
\label{fig:Fig1}
\end{figure}

The adiabatic generalization of the Floquet-wave (FW) theorem is a well known method for the analysis of leaky-wave holograms generating pencil beams in the microwave regime, which was initially proposed by Minatti et al.  \cite{minatti_2016}.
The advantage of using this method is the precise analysis of anisotropic holograms without the need for full-wave simulations, which significantly reduces the synthesis time consumption.
A conceptual structure of anisotropic hologram is shown in Fig.\ref{fig:Fig1a}, including the  modulated metasurface (hologram) and vertical monopole as surface wave generator. 
The monopole excites a cylindrical magnetic surface wave, which can be represented by \cite{moeini_scirep_2019}:
\begin{equation}
\vec{H}_t|_{z=0^+}\approx \vec{J}_{sw}H_1^{(2)}(k^{(0)}\rho)
\end{equation}
where $H_1^{(2)}$ is the Hankel function of second kind and first order. Also,  $k^{(0)}$ is the complex surface wave number. 
The metasurface consists of pseudo-periodic patches printed on the grounded dielectric substrate (as shown in Fig.\ref{fig:Fig1b}).  If the dimensions of patches are small enough compared to the wavelength ($\approx\lambda/6$), the metasurface effectively acts as an impedance surface. By proper patterning of surface impedance, the surface wave can be converted to leaky mode with the desired direction, polarization and topological charge. The main purpose of using adiabatic Floquet-wave analysis is to provide a systematic approach for calculating the distribution of leakage parameter ($\alpha$) across the aperture, that can be used to synthesize radiation patterns with the desired specifications \cite{minatti_2_2016}. In this paper, this method is used for the analysis of anisotropic holograms with the ability to generate vortex waves carrying OAM modes.
\subsection{Surface impedance distribution}
In the analysis of leaky-wave holograms, the characteristics of metasurface can be described by the impedance boundary condition. For the metasurface placed at z = 0 (see Fig.\ref{fig:Fig1b}), the transparent impedance boundary condition can be expressed as \cite{tretyakov_2003}:
\begin{equation}
\vec{E}_t = j\bar{\bar{X}}(\vec{\rho}).\hat{z}(\vec{H}_t|_{z=0^+} - \vec{H}_t|_{z=0^-})
 = j\bar{\bar{X}}(\vec{\rho}).\vec{J}
 \label{eq:Et}
\end{equation}
where
\begin{equation}
\bar{\bar{X}}(\vec{\rho}) = 
\begin{pmatrix}
X_{\rho\rho}(\vec{\rho}) & X_{\rho\phi}(\vec{\rho}) \\
X_{\rho\phi}(\vec{\rho}) & X_{\phi\phi}(\vec{\rho})
\end{pmatrix}
\end{equation}
Note that, $\bar{\bar{X}}(\vec{\rho})$ indicates the tensorial reactance and is dependent on the observation vector $\vec{\rho}$ in the cylindrical coordinates.
$\vec{E}_t$ and $\vec{H}_t$ represent the tangential electric and magnetic fields, respectively.
 In the general case, the distribution of surface reactance can be expressed as follows, which are obtained directly from the generalized holographic theory \cite{minatti_2016}:
\begin{equation}
X_{\rho\rho}(\vec{\rho})=X_\rho [1+m_\rho(\vec{\rho})\cos (Ks(\vec{\rho}) + \Phi_{\rho}(\vec{\rho}))]
\label{eq:X_rho_rho}
\end{equation}
\begin{equation}
X_{\rho\phi}(\vec{\rho})=X_\rho m_\phi(\vec{\rho})\cos(Ks(\vec{\rho}) + \Phi_\phi(\vec{\rho}))
\label{eq:X_rho_phi}
\end{equation}
\begin{equation}
X_{\phi\phi}(\vec{\rho})=X_\phi[1-m_\rho(\vec{\rho})\cos(Ks(\vec{\rho}) + \Phi_\rho(\vec{\rho}))]
\label{eq:X_phi_phi}
\end{equation}
In (\ref{eq:X_rho_rho})-(\ref{eq:X_phi_phi}) the coefficients $X_\rho$ and $X_\phi$ demonstrate the average surface reactances, which are independent from the position vector $\vec{\rho}$. Also, $m_\rho$ and $m_\phi$ are the modulation indices controlling the distribution of leakage constant across the radiation aperture. Note that, $Ks(\vec{\rho})$ is the rapidly varying part and $\Phi_{\rho,\phi}(\vec{\rho})$ are the slowly varying parts of modulation phase such that:
\begin{equation}
|\nabla_{\vec{\rho}}Ks(\vec{\rho})|\gg|\nabla_{\vec{\rho}}\Phi_{\rho, \phi}(\vec{\rho})|
\label{eq:nabla}
\end{equation}
where $\nabla_{\vec{\rho}}$ is the gradient operator on $\vec{\rho}$.
The slowly varying parts of modulation phase ($\Phi_{\rho, \phi}(\vec{\rho})$) play an important role in control of polarization and beam vorticity. If they are linear functions of azimuth angle ($\phi$) such that:
\begin{equation}
\Phi_{\rho}(\vec{\rho}) = \Phi_{\phi}(\vec{\rho})=\pm l\phi   \quad   l = 0, 1, 2, ... 
\end{equation}
then the radiated beam can support OAM mode with an integer topological charge. Note that, the local periodicity of the boundary condition in the $\rho$ direction is obtained by:
\begin{equation}
p(\vec{\rho}) =\frac{2\pi}{\nabla_{\vec{\rho}}(Ks(\vec{\rho})\pm l\phi).\hat{\rho}} = \frac{2\pi}{\nabla_{\vec{\rho}}(Ks(\vec{\rho}).\hat{\rho}} 
\end{equation}
In order to determine $Ks$, $\Phi_\rho$ and $\Phi_\phi$, we will use the AFE method \cite{casaletti_2017, amini_2020}. In this method, the aperture field is estimated from the inverse Fourier transformation of the desired radiation pattern.
\subsection{Floquet-wave expansion of surface current}
According to the periodicity of surface impedance in (\ref{eq:X_rho_rho})-(\ref{eq:X_phi_phi}), the higher order modes existing in the surface current must be considered. This suggests that the induced surface current may be expanded in terms of $n$'th harmonic as follows \cite{oliner_1959, minatti_2016}:
\begin{equation}
\vec{J}\approx \sum _{n=-\infty}^{+\infty}(J_\rho^{(n)}\hat{\rho}  +  J_\phi^{(n)}\hat{\phi})e^{-jnKs(\vec{\rho})}H_1^{(2)}(k^{(0)}\rho)
\label{eq:J}
\end{equation}
Using asymptotic form of the Hankel function, the equation (\ref{eq:J}) can be written as:
\begin{equation}
\vec{J}\approx \sum _{n=-\infty}^{+\infty} \sqrt{\frac{2j}{\pi k^{(0)} \rho}}(J_\rho^{(n)}\hat{\rho}  +  J_\phi^{(n)}\hat{\phi})e^{-j(nKs(\vec{\rho}) + k^{(0)}\rho)}
\label{eq:J_mod}
\end{equation} 
Note that the spatial derivative of the phase in (\ref{eq:J_mod}) gives the $n$-indexed complex wave vector:
\begin{equation}
\vec{k}^{(n)}(\vec{\rho}) = \vec{\beta}^{(n)}(\vec{\rho}) - j\delta\vec{\alpha}(\vec{\rho}) =  \nabla_{\vec{\rho}}(k^{(0)}\rho + nKs(\vec{\rho}))
\end{equation}
where $\delta\vec{\alpha}$ is the leakage vector and represents the energy converted from surface wave into leaky wave and $\vec{\beta}^{(n)}$ is the propagation vector corresponding to the $n$'th harmonic of Floquet mode. Given that the value of $Ks(\vec{\rho})$ is assumed to be real, the leakage vector is independent of $n$. So we can write:
\begin{equation}
\begin{split}
\vec{\beta}^{(n)}(\vec{\rho})= Re\{\vec{k}^{(n)}(\vec{\rho})\} = Re\{\nabla_{\vec{\rho}}(k^{(0)}\rho+nKs(\vec{\rho}))\}=\\
\vec{\beta}^{(0)}_{um} + \delta \vec{\beta} (\vec{\rho}) + n\nabla_{\vec{\rho}}Ks(\vec{\rho})\\
n = 0, \pm1, \pm2, ...
\end{split}
\label{eq:beta_n}
\end{equation}
\begin{equation}
\delta \vec{\alpha}(\vec{\rho})=-Im\{\vec{k}^{(n)}(\vec{\rho})\} = -Im\{\nabla_{\vec{\rho}}(k^{(0)}\rho)\}
\label{eq:alpha_n}
\end{equation}
In (\ref{eq:beta_n}), $\vec{\beta}_{um}^{(0)}$ is the propagation vector for unmodulated impedance (i.e. $m_\rho(\vec{\rho})=m_\phi(\vec{\rho})=0$). In this case the wave is confined on the surface and the wave number becomes purely real. The vectors
$\delta \vec{\beta}$ and $\delta \vec{\alpha}$ are the small deviations with respect to the $\vec{\beta}_{um}^{(0)}$ when  modulation is applied.
If the parameters $m_\rho$ and $m_\phi$ have small values ($|m_{\rho,\phi}|\leq1$), we can neglect $\delta \vec{\beta}$ in the calculation of propagation vector \cite{oliner_1959, patel_2011}.
To have a comprehensive insight of the radiation mechanism, the parameters $\vec{\beta}_{um}^{(0)}$ and $\delta \vec{\alpha}$ must be accurately determined. 
For this purpose, the expansion of electrical field in (\ref{eq:Et}) must be written in terms of Floquet waves and an Eigen-value problem should be solved.
By considering sufficiently large numbers of Floquet modes, the desired accuracy can be achieved. 
As an alternative solution,
in \cite{minatti_2016}  closed-form expressions are derived for $\vec{\beta}_{um}^{(0)}$ and $\delta\vec{\alpha}$ which are sufficiently accurate for our purpose. 
In this paper, the latter is used to determine the complex wave vector.
Using transverse resonance technique for TM-like modes,  yields
\begin{equation}
\vec{\beta}_{um}^{(0)}=\hat{\rho}k\sqrt{1+(\frac{X_s}{\eta_0})^2}
\end{equation}
where $X_s$ is obtained from the following nonlinear equation:
\begin{equation}
X_s = X_\rho[1-\frac{X_s \epsilon_r \cot(kh\sqrt{\epsilon_r - 1 - (X_s/\eta_0)^2})}{\eta_0 \sqrt{\epsilon_r - 1 - (X_s/\eta_0)^2}}]
\end{equation}
For $\delta \vec{\alpha}$ we have:
\begin{equation}
\delta \vec{\alpha}(\vec{\rho})\approx\hat{\rho}\frac{-Re\{(\hat{\rho}-\frac{\chi_{\phi\rho}^*}{\chi_{\phi\phi}^*}\hat{\phi}).\bar{\bar{z}}^{(-1)\dagger}.(\hat{\rho}-\frac{\chi_{\phi\rho}}{\chi_{\phi\phi}}\hat{\phi})\}}{\frac{X_\rho^2\beta_{um}^{(0)}k}{\eta_0}[\frac{2\epsilon_r}{h(\epsilon_r k^2 - (\beta_{um}^{(0)})^2)^2}   +   \frac{1}{((\beta_{um}^{(0)})^2 - k^2)^{3/2}}]}
\label{eq:aalpha}
\end{equation} 
where $\chi_{\phi\rho}$ and $\chi_{\phi\phi}$ are the components of the tensor $\bar{\bar{\chi}}$ (see Appendix A). Both $\bar{\bar{\chi}}$ and $\bar{\bar{z}}$ are dependent on the dyadic Green's function of grounded slab. The process of calculating the quantities in (\ref{eq:aalpha}) is described in Appendix A. 
In Fig.\ref{fig:delta_alpha}, the colored map of $\delta\alpha/k$ in terms of modulation indices (for the case of $m_\rho = m_\phi$) and $X_s$ is plotted. Rogers RO4003 with $\epsilon_r = 3.55$ and $h=1.524mm$ is chosen as the dielectric host medium. Observe that for $X_s < 0.8\eta_0$ ($\eta_0$  as the free-space impedance), with increasing modulation index, the leakage constant will increase and would vary from 0 to $0.015k$. For $X_s>0.8\eta_0$ the leakage constant peaks at a certain value and decreases again thereafter.
\begin{figure}
\includegraphics[width = 0.37\textwidth]{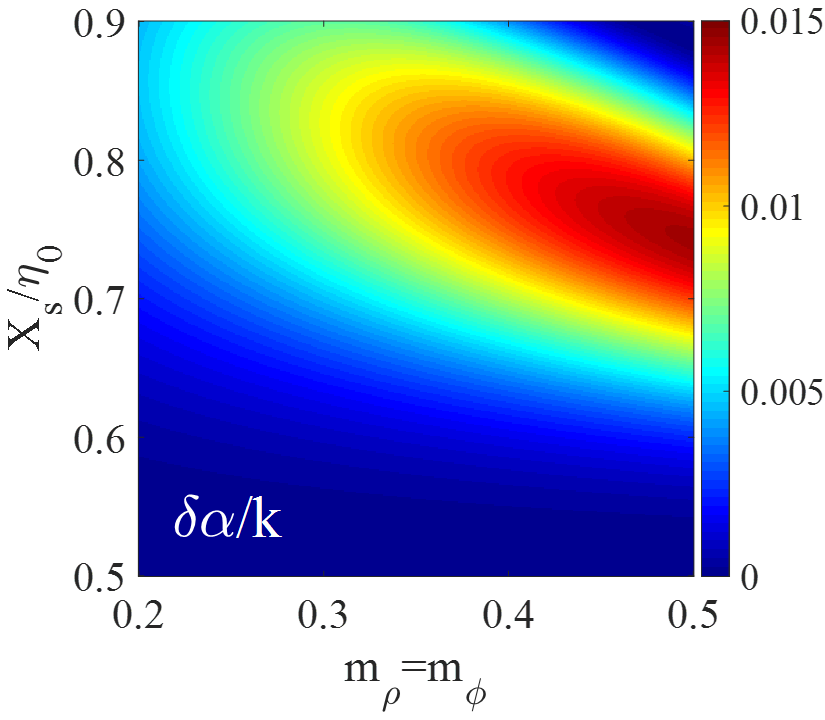}
\caption{Variation of $\delta\alpha$ versus modulation index and $X_s$.}
\label{fig:delta_alpha}
\end{figure}
\subsection{Estimation of $Ks(\vec{\rho})$ and $\Phi_{\rho,\phi}(\vec{\rho})$}
In this section, we develop the theoretical procedure for the synthesis of vortex wave with arbitrary topological charge using leaky-wave holograms. 
A common method for designing leaky-wave holograms is to use the aperture field estimation technique \cite{casaletti_2017, minatti_2015}. In this method the relationship between the surface reactance tensor and the aperture field vector ($\vec{E}_{ap}$) can be expressed as \cite{minatti_2015}:
\begin{equation}
\bar{\bar{X}}(\vec{\rho}).\hat{\rho} = X_0[\hat{\rho} + 2Im\{\frac{\vec{E_{ap}}}{-J_{sw}H_1^{(2)}(k^{(0)}\rho)}\}]
\label{eq:Xs_Ea}
\end{equation}
where $H_1^{(2)}$ denotes the surface wave function excited by the monopole launcher. 
Using the asymptotic expansion of Hankel function in (\ref{eq:Xs_Ea}) yields
\begin{equation}
\begin{split}
\bar{\bar{X}}(\vec{\rho}).\hat{\rho}=X_0[\hat{\rho} + \frac{\sqrt{2\pi\rho\sqrt{(\beta_{um}^{(0)} + \delta\beta (\vec{\rho}))^2 + \delta\alpha^2(\vec{\rho})}}}{J_{sw}}\times\\
Im \{\vec{E}_{ap}je^{j(k^{(0)}\rho-\zeta_0)}\}]
\end{split}
\label{eq:Xs_asym}
\end{equation}
The aperture field vector can be expanded in terms of x and y components:
\begin{equation}
\vec{E}_{ap}(\vec{\rho})=E_{ax}(\vec{\rho})\hat{x}+E_{ay}(\vec{\rho})\hat{y}
\label{eq:Eap}
\end{equation}
To estimate the aperture field vector, three points must be considered: 1- The distribution of field magnitude that controls the shape of radiated beam; 2- The phase of field that determines the direction and vorticity state of beam; 3- The relationship between the x and y components of the aperture field that determines the polarization state of radiated wave.

In order to have an object wave propagating in $\theta_0$ and $\phi_0$ direction and carrying OAM mode with arbitrary topological charge, we can define the components of aperture field vector as
\begin{equation}
\begin{split}
E_{ax, ay}(\vec{\rho})=M_{x, y}(\vec{\rho})\frac{J_{sw}}{\sqrt{2\pi\rho\sqrt{(\beta_{um}^{(0)} + \delta\beta (\vec{\rho}))^2 + \delta\alpha^2(\vec{\rho})}}}\times\\
e^{-\delta\alpha (\vec{\rho})\rho} e^{-j(k\rho \sin \theta_0 \cos (\phi - \phi_0)+ l\phi)}
\end{split}
\label{eq:Eaxay}
\end{equation} 
Note that, $M_x$ and $M_y$ may be dependent on $\vec{\rho}$. To obtain a helical wave front , the term of $l\phi$ is added to the phase of aperture field.
In (\ref{eq:Eaxay}), for simplicity, the field amplitude is chosen so that only $M_x$ and $M_y$ appear in the impedance equation (Eq. (\ref{eq:Xs_asym})). 
Substituting  (\ref{eq:Eaxay}) in (\ref{eq:Xs_asym}) and comparing the obtained impedance function with (\ref{eq:X_rho_rho})-(\ref{eq:X_phi_phi}), yields
\begin{equation}
Ks(\vec{\rho})=(\beta^{(0)}_{um} + \delta  \beta(\vec{\rho}))\rho - k \sin \theta_0 \cos (\phi-\phi_0)\rho
\end{equation}
\begin{equation}
\Phi_\rho(\vec{\rho})=l\phi - arg\{M_x(\vec{\rho})\cos \phi + M_y(\vec{\rho})\sin \phi\}
\label{eq:Phi_rho}
\end{equation}
\begin{equation}
\Phi_\phi(\vec{\rho})=l\phi - arg\{-M_x(\vec{\rho})\sin \phi + M_y(\vec{\rho})\cos \phi\}
\label{eq:Phi_phi}
\end{equation}
\begin{equation}
m_\rho(\vec{\rho}) = |M_x(\vec{\rho})\cos \phi + M_y(\vec{\rho})\sin \phi|
\label{eq:m_rho}
\end{equation}
\begin{equation}
m_\phi(\vec{\rho})=|-M_x(\vec{\rho})\sin \phi + M_y(\vec{\rho})\cos \phi|
\label{eq:m_phi}
\end{equation}
Using (\ref{eq:beta_n}) the propagation constant for $n=-1$ (radiation mode) can be obtained as
\begin{equation}
\begin{split}
\vec{\beta}^{(-1)}(\vec{\rho})=(\beta_{um}^{(0)}+\delta \beta(\vec{\rho})\hat{\rho} - \nabla_{\vec{\rho}}ks(\vec{\rho}) = \\
k\sin \theta_0 \cos (\phi - \phi_0)\hat{\rho} - 
k\sin \theta_0 \sin (\phi - \phi_0)\hat{\phi}
\end{split}
\end{equation}
Therefore, $\beta^{(-1)}$ always satisfies the radiation condition, that is
\begin{equation}
|\vec{\beta}^{(-1)}(\vec{\rho})|=k\sin \theta_0<k
\end{equation}
\section{Calculation of far-zone field}
In the analysis of radiative apertures, the Fourier transformation and stationary phase theory  can be applied to determine the far-zone fields. Thus, for the $\theta$ and $\phi$ components of the far-zone field we have \cite{balanis_2016}:
\begin{equation}
F_\theta(\theta, \phi) = \tilde{E}_{ax} \cos \phi + \tilde{E}_{ay} \sin \phi
\label{eq:FF_theta}
\end{equation}
\begin{equation}
F_\phi (\theta, \phi) = \cos \theta (-\tilde{E}_{ax} \sin \phi + \tilde{E}_{ay} \cos \phi)
\label{eq:FF_phi}
\end{equation}
where $\tilde{E}_{ax}$ and $\tilde{E}_{ay}$ are the Fourier transforms of $E_{ax}$ and $E_{ay}$ respectively. Hence, in the cylindrical coordinate system we can write
\begin{equation}
\tilde{E}_{ax} = \iint_{ap}E_{ax}e^{jk\rho'\sin \theta \cos (\phi - \phi')}\rho'd\rho' d\phi'
\end{equation} 
\begin{equation}
\tilde{E}_{ay} = \iint_{ap}E_{ay}e^{jk\rho'\sin \theta \cos (\phi - \phi')}\rho'd\rho' d\phi'
\end{equation} 
\section{Design and analysis of hologram}
\begin{figure}
\includegraphics[width = 0.3\textwidth]{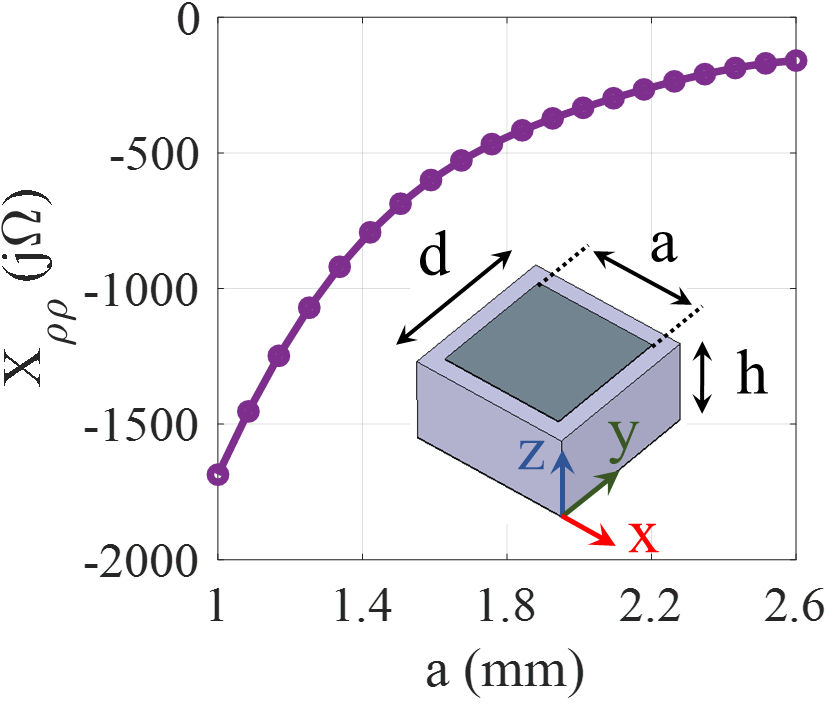}
\caption{Variation of scalar surface reactance for square patch.}
\label{fig:square_patch}
\end{figure}

To validate the analytical approach in the previous section, we present two examples of leaky-wave holograms generating OAM modes. In the first section, an isotropic unit cell is used to realize hologram with a broadside beam.
In the second, anisotropic hologram with circular polarization is designed. Without loss of generality, the operation frequency of antennas is selected to be 18 GHz. 
Rogers RO4003 with dielectric constant  3.55, loss tangent $\tan\delta = 0.0027$ and thickness  1.524 mm is used as the substrate. 
The square patch printed on the grounded dielectric has  been chosen for realization of isotropic (scalar) impedance. The period of unit cell is chosen as  2.8 mm which is equal to $\lambda/6$ at the operating frequency. Fig.\ref{fig:square_patch} shows the proposed unit cell and reactance curve for variation of square width (namely $a$), which is obtained by Eigen-mode solver in CST microwave studio \cite{cst}. 

\begin{figure}
 \centering
 \begin{subfigure}[b]{0.3\textwidth}
		\includegraphics[width =\textwidth]{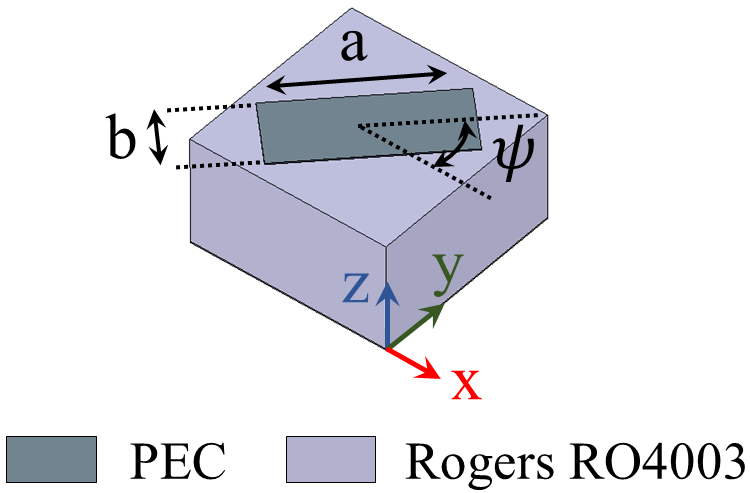}
		\caption{}
		\label{fig:UC_aniso_a}
\end{subfigure}
 \begin{subfigure}[b]{0.235\textwidth}
		\includegraphics[width =\textwidth]{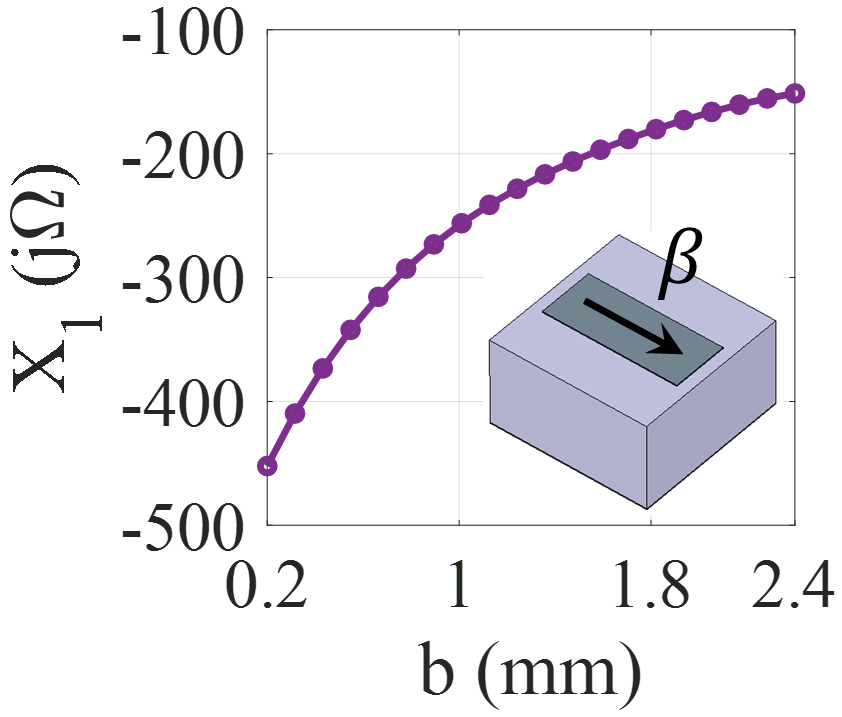}
		\caption{}
		\label{fig:UC_aniso_b}
\end{subfigure}
\begin{subfigure}[b]{0.235\textwidth}
		\includegraphics[width =\textwidth]{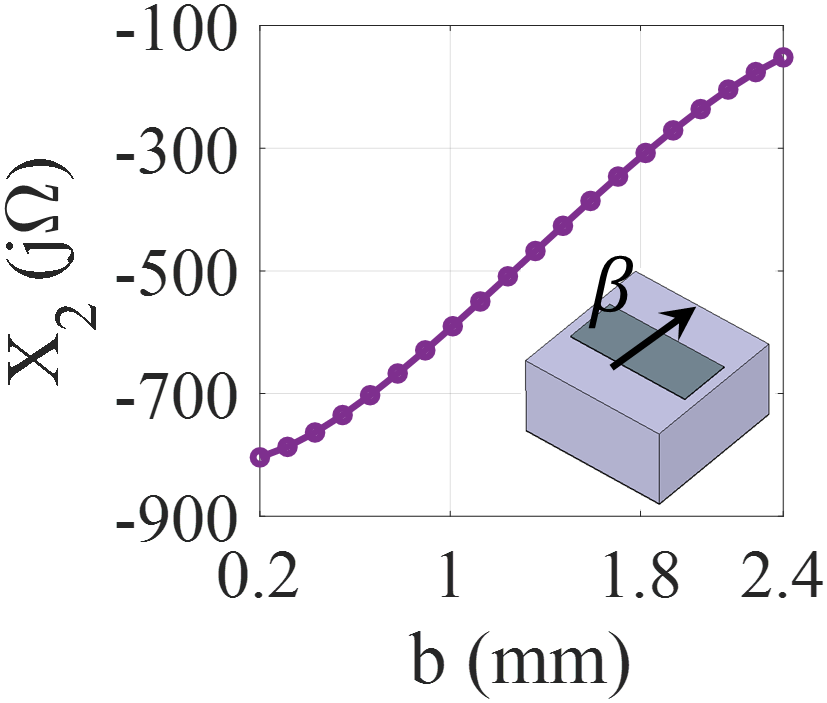}
		\caption{}
		\label{fig:UC_aniso_c}
\end{subfigure}
\caption{(a) Asymmetric rectangular patch for realization of tensorial impedance. (b) Impedance curve for the wave propagating along the x axis  ($X_1$). (c) Impedance curve for the wave propagating along the y axis ($X_2$).}
\label{fig:UC_aniso}
\end{figure}

\begin{figure}
\begin{subfigure}[b]{0.23\textwidth}
		\includegraphics[width =\textwidth]{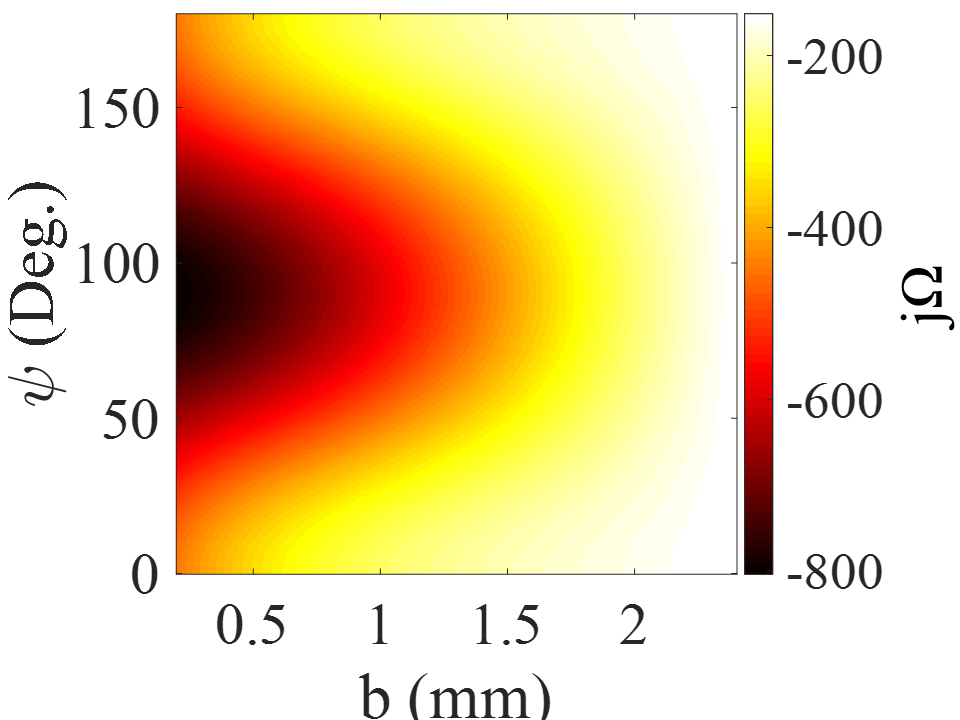}
		\caption{}
\end{subfigure}
\begin{subfigure}[b]{0.23\textwidth}
		\includegraphics[width =\textwidth]{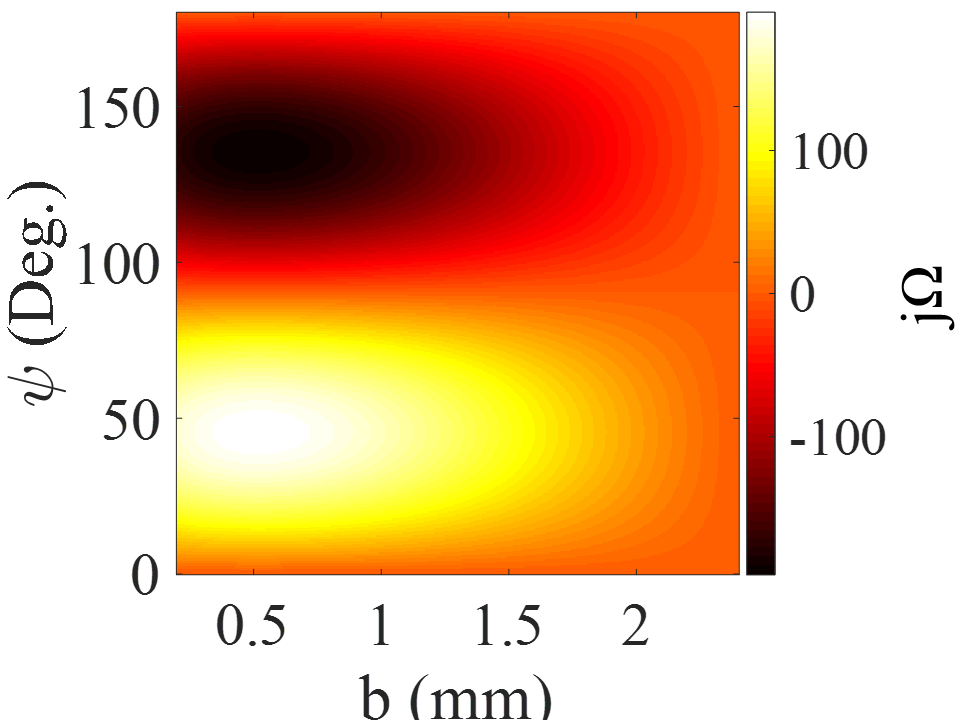}
		\caption{}
\end{subfigure}
\caption{Design maps for asymmetric rectangular patch in Fig.\ref{fig:UC_aniso_a} in terms of width and  orientation angle. (a) $X_{\rho\rho}$. (b) $X_{\rho\phi}$.}
\label{fig:psi_b}
\end{figure}

 To implement the anisotropic (tensorial) impedances,  we utilize asymmetric rectangular patches as the composition pixels. Fig.\ref{fig:UC_aniso_a} shows the proposed anisotropic unit cell. The width of the patch (namely $b$) and its orientation angle (namely $\psi$) are considered for changing the impedance. 
To extract the impedance tensor, first the reactances $X_1$ and $X_2$ are calculated for the surface wave propagating along the x and y directions, respectively. In both directions, the parameter $\psi$ is kept equal to zero and the parameter $b$ is varied from 0.2 to 2.4 mm. 
The reactance curves for $X_1$ and $X_2$ are plotted in Figs \ref{fig:UC_aniso_b} and \ref{fig:UC_aniso_c}, respectively. 
 For the retrieval of the impedance tensor the method proposed in \cite{werner_2014} is used, which can be expressed as
 \begin{equation}
\bar{\bar{Z}}_s=j\bar{\bar{X}}_s=jR^T(\psi)\bar{\bar{X}}(b)R(\psi)
\end{equation}
where
\begin{equation}
\bar{\bar{X}}(b) = 
\begin{pmatrix}
X_1(b) & 0 \\
0 & X_2(b)
\end{pmatrix}
\end{equation}
and $R$ is the rotation matrix:
\begin{equation}
R(\psi) = 
\begin{pmatrix}
\cos \psi & -\sin \psi\\
\sin \psi & \cos \psi
\end{pmatrix}
\end{equation}
Note that $T$ indicates the transpose operator.
Fig.\ref{fig:psi_b} shows the impedance maps of proposed anisotropic unit cell versus the width and orientation angle of patch.
\subsection{Aperture field for isotropic holograms}
\begin{figure}
\includegraphics[width = 0.44\textwidth]{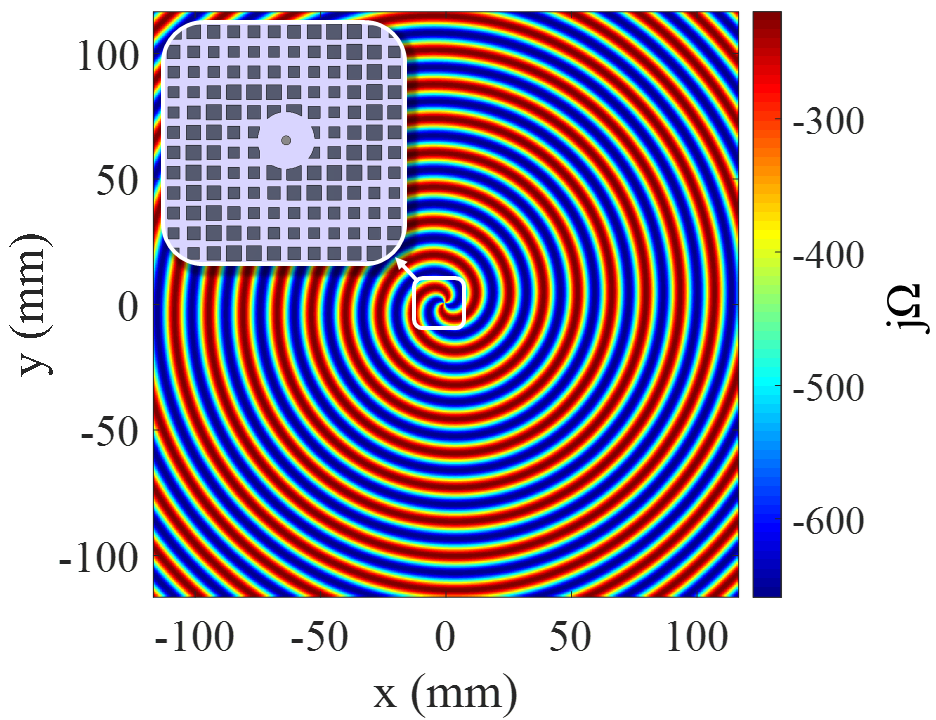}
\caption{Scalar impedance pattern for generating OAM wave with topological charge of m = 2.}
\label{fig:imp_iso}
\end{figure}

For the isotropic surface reactance, the aperture field should have only the $\rho$ component in (\ref{eq:Xs_Ea}). In this case, the x and y components of the aperture field cannot be defined independently and will be related as follows:
\begin{equation}
E_{ax}(\vec{\rho})\sin \phi = E_{ay}(\vec{\rho})\cos \phi
\end{equation}
Isotropic conditions impose some constraints on the definition of the field components. Due to this limitation, polarization control of leaky modes on isotropic holograms is difficult.
According to the above considerations, the x and y components of the field are selected  as:
\begin{equation}
\begin{split}
E_{ax}(\vec{\rho})=M\cos \phi\frac{J_{sw}}{\sqrt{2\pi\rho\sqrt{(\beta_{um}^{(0)} + \delta\beta (\vec{\rho}))^2 + \delta\alpha^2(\vec{\rho})}}}\times\\
e^{-\delta\alpha (\vec{\rho})\rho} e^{-j(k\rho \sin \theta_0 \cos (\phi - \phi_0)+ l\phi)}
\end{split}
\label{eq:Mx_iso}
\end{equation}
\begin{equation}
\begin{split}
E_{ay}(\vec{\rho})=M\sin \phi\frac{J_{sw}}{\sqrt{2\pi\rho\sqrt{(\beta_{um}^{(0)} + \delta\beta (\vec{\rho}))^2 + \delta\alpha^2(\vec{\rho})}}}\times\\
e^{-\delta\alpha (\vec{\rho})\rho} e^{-j(k\rho \sin \theta_0 \cos (\phi - \phi_0)+ l\phi)}
\end{split}
\label{eq:My_iso}
\end{equation}
In (\ref{eq:Mx_iso}) and (\ref{eq:My_iso}), the modulation indices $M_x$ and $M_y$ are selected as $M \cos \phi$ and $M \sin \phi$, respectively. Substituting (\ref{eq:Mx_iso}) and (\ref{eq:My_iso}) in (\ref{eq:m_rho}) and (\ref{eq:m_phi}) concludes that:
\begin{equation}
m_\rho (\vec{\rho}) = M
\end{equation}
\begin{equation}
m_\phi (\vec{\rho}) = 0
\end{equation}
which indicates that $X_{\rho\phi}$ is zero.
Also
\begin{equation}
\Phi_\rho (\vec{\rho}) = \Phi_\phi(\vec{\rho}) = l \phi
\end{equation}
Without loss of generality, if the hologram is designed to radiate vortex wave in the direction of broadside (namely, $\theta_0 = 0^\circ$ and $\phi_0 = 0^\circ$), by substituting (\ref{eq:Mx_iso}) and (\ref{eq:My_iso}) in (\ref{eq:FF_theta}) and (\ref{eq:FF_phi}) the far-zone fields can obtained as:
\begin{equation}
\begin{split}
F_\theta(\theta, \phi) = \iint_{ap}\frac{ M \cos (\phi-\phi') J_{sw}}{\sqrt{2\pi\rho \sqrt{(\beta_{um}^{(0)} + \delta \beta(\vec{\rho}))^2 + \delta \alpha^2 (\vec{\rho})}}}\times\\
e^{-\delta \alpha (\vec{\rho})} e^{jk\rho' \sin \theta \cos (\phi - \phi') } e^{-jl\phi'}\rho'd\rho'd\phi'
\end{split}
\label{eq:int_iso1}
\end{equation}
\begin{equation}
\begin{split}
F_\phi(\theta, \phi) = \iint_{ap}\frac{ -M \sin (\phi-\phi') \cos \theta J_{sw}}{\sqrt{2\pi\rho \sqrt{(\beta_{um}^{(0)} + \delta \beta(\vec{\rho}))^2 + \delta \alpha^2 (\vec{\rho})}}}\times\\
e^{-\delta \alpha (\vec{\rho})} e^{jk\rho' \sin \theta \cos (\phi - \phi') } e^{-jl\phi'}\rho'd\rho'd\phi'
\end{split}
\label{eq:int_iso2}
\end{equation}
Setting $u = \phi - \phi'$ and using substitution rule in (\ref{eq:int_iso1}) and (\ref{eq:int_iso2}), we can conclude that the phases of $F_\theta$ and $F_\phi$ have the angular dependence in the form of $l\phi$. This means that the number of twists of the resulting wave front around the singularity point is $m = l$, which indicates the topological charge of radiated wave \cite{vaity_2015}.
To investigate the validity of the proposed method, here, an isotropic hologram with the ability to generate vortex beam with topological charge of m = 2 has been designed. The direction of object wave is supposed to be broadside ($\theta_0 = 0$).
Fig.\ref{fig:imp_iso} shows the scalar surface impedance pattern ($X_{\rho\rho}$) and realized model for $M=0.5$ and $X_{s}=0.7\eta_0$. The parameters $M$ and $X_{s}$ are selected such that the calculated impedances can be realized easily. The overall dimensions of the hologram are large enough ($14\lambda\times 14\lambda$) for the surface wave to be effectively converted into leaky wave.
 \begin{figure*}
\begin{subfigure}{0.24\textwidth}
		\includegraphics[width =\textwidth]{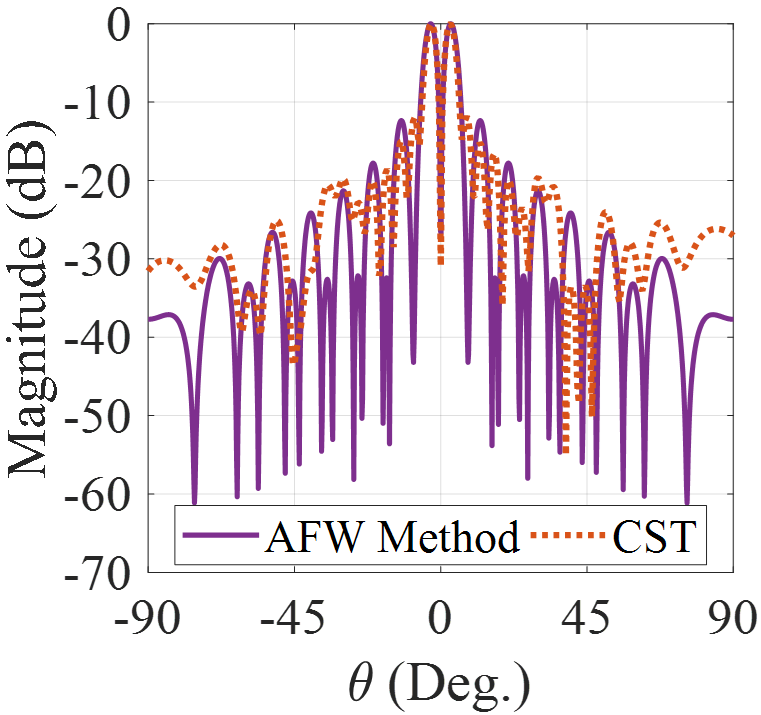}
		\caption{}
\end{subfigure}
\begin{subfigure}{0.24\textwidth}
		\includegraphics[width =\textwidth]{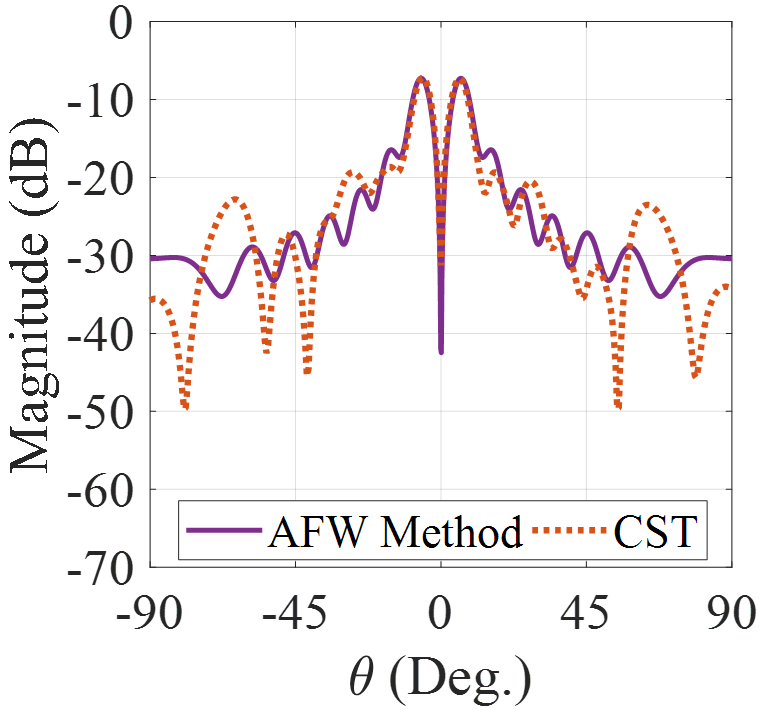}
		\caption{}
\end{subfigure}
\begin{subfigure}{0.24\textwidth}
		\includegraphics[width =\textwidth]{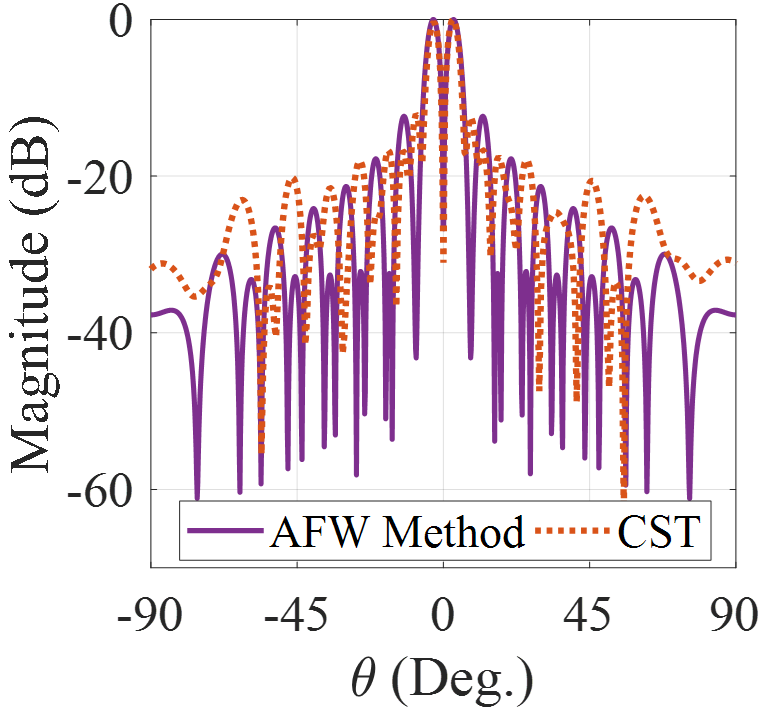}
		\caption{}
\end{subfigure}
\begin{subfigure}{0.24\textwidth}
		\includegraphics[width =\textwidth]{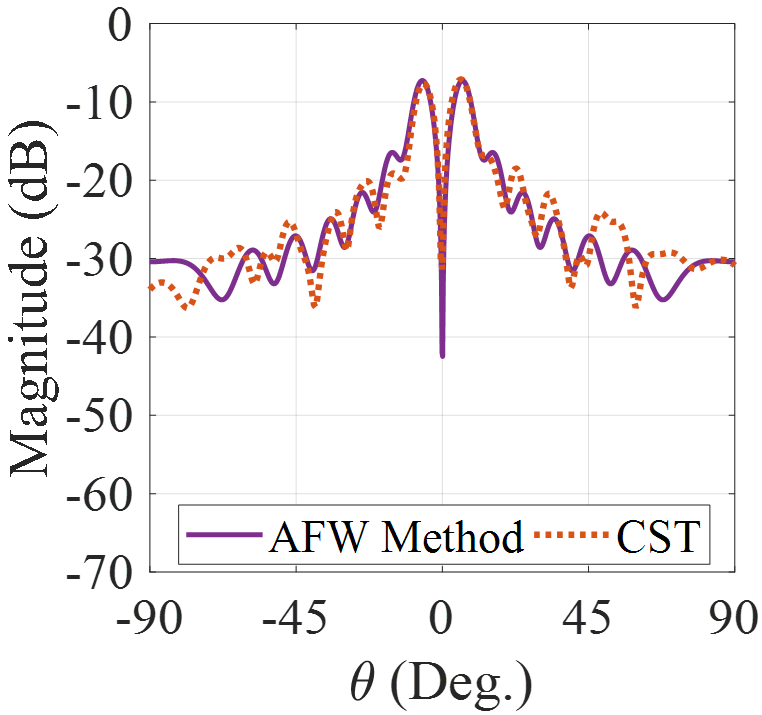}
		\caption{}
\end{subfigure}
\caption{Comparison between the analytical and simulation results of isotropic hologram. (a) RHCP component at $\phi=0^\circ$. (b) LHCP component at $\phi=0^\circ$. (c) RHCP component at $\phi=90^\circ$. (d) LHCP component at $\phi=90^\circ$. }
\label{fig:iso_patterns}
\end{figure*}
 In Fig.\ref{fig:iso_patterns} the results of analytical model and full-wave simulation for radiation patterns are compared, which are in good agreement.
 Note that although there is no degree of freedom to control the polarization of aperture field,  the main beam is circularly polarized, which is caused by spiral variations in the scalar surface impedance (see Fig.\ref{fig:imp_iso}) \cite{minatti_2011}. Results in Fig.\ref{fig:iso_patterns} show that the LHCP component (cross-pol.) of far-field pattern is approximately 8 dB lower than RHCP component (co-pol.).
Fig.\ref{fig:phase_iso} shows the phase distributions of $F_\theta$ and $F_\phi$. Observe that the number of twists of wave front is 2 and the phase singularity occurs at the broadside. Note that the phase of $F_\theta$ has a desirable form only at a very small spatial angle. This issue will be solved by anisotropic structures.
\begin{figure}
\begin{subfigure}[b]{0.237\textwidth}
		\includegraphics[width =\textwidth]{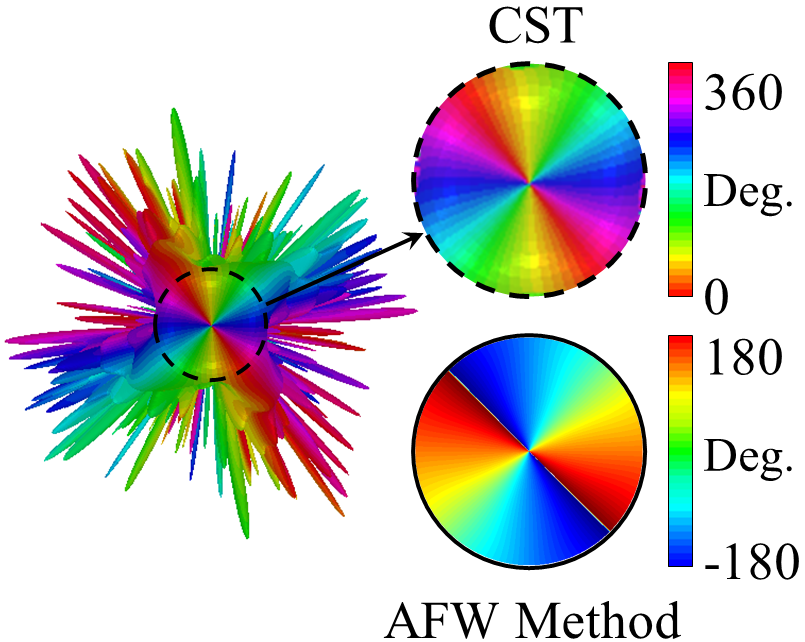}
		\caption{}
\end{subfigure}
\begin{subfigure}[b]{0.237\textwidth}
		\includegraphics[width =\textwidth]{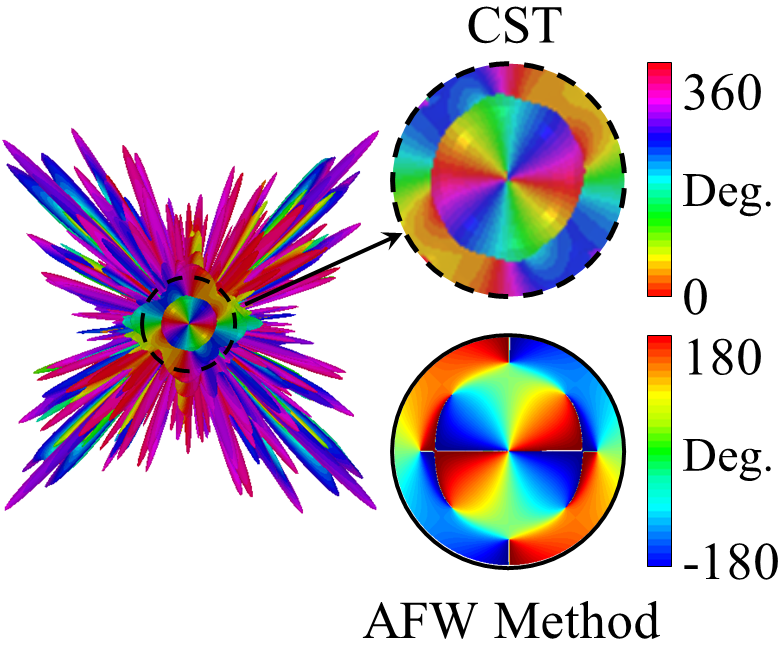}
		\caption{}
\end{subfigure}
\caption{Phase distributions of far-zone components. (a) $F_\phi$.  (b) $F_\theta$, for isotropic hologram in Fig.\ref{fig:imp_iso}.  }
\label{fig:phase_iso}
\end{figure} 
\subsection{Aperture field for anisotropic holograms}
\begin{figure*}[t]
\begin{subfigure}[b]{0.5\textwidth}
	\includegraphics[width = \textwidth]{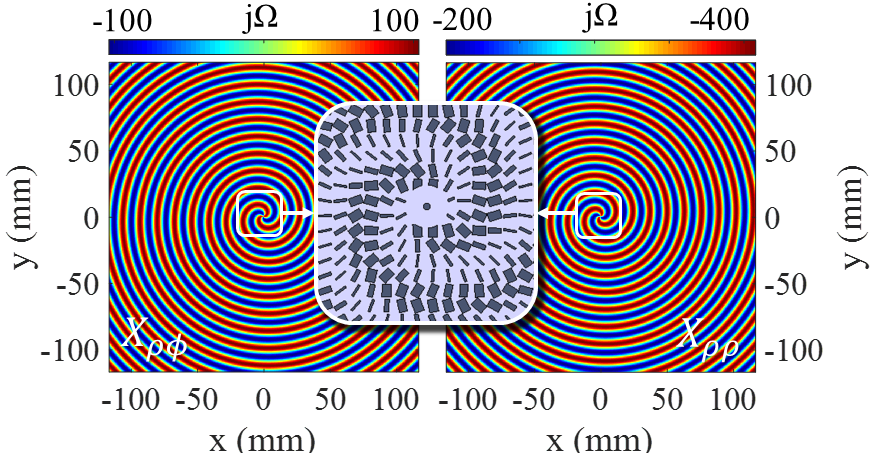}
	\caption{}
	\label{fig:imp_aniso}
\end{subfigure}
\begin{subfigure}[b]{0.22\textwidth}
		\includegraphics[width =\textwidth]{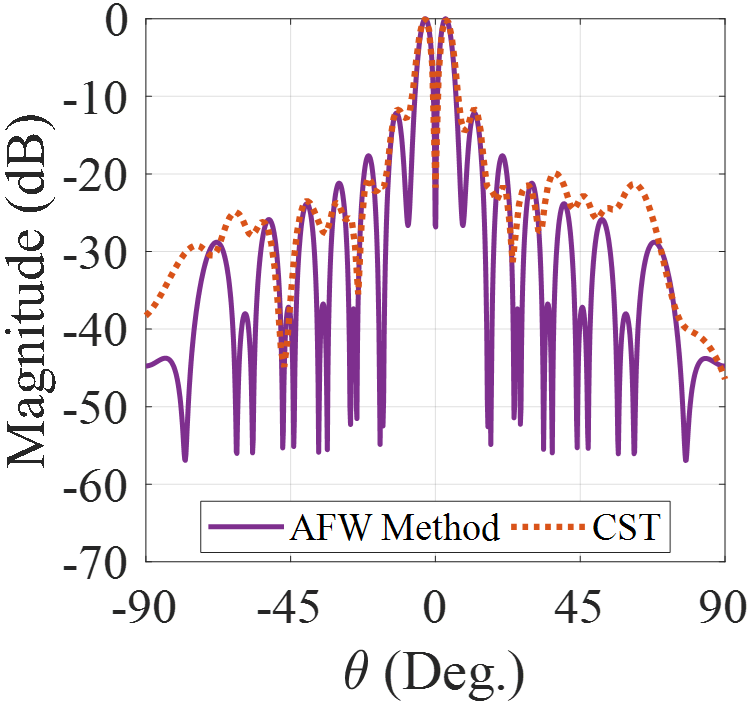}
		\caption{}
		\label{fig:pattern_comp_aniso_phi0}
\end{subfigure}
\begin{subfigure}[b]{0.22\textwidth}
		\includegraphics[width =\textwidth]{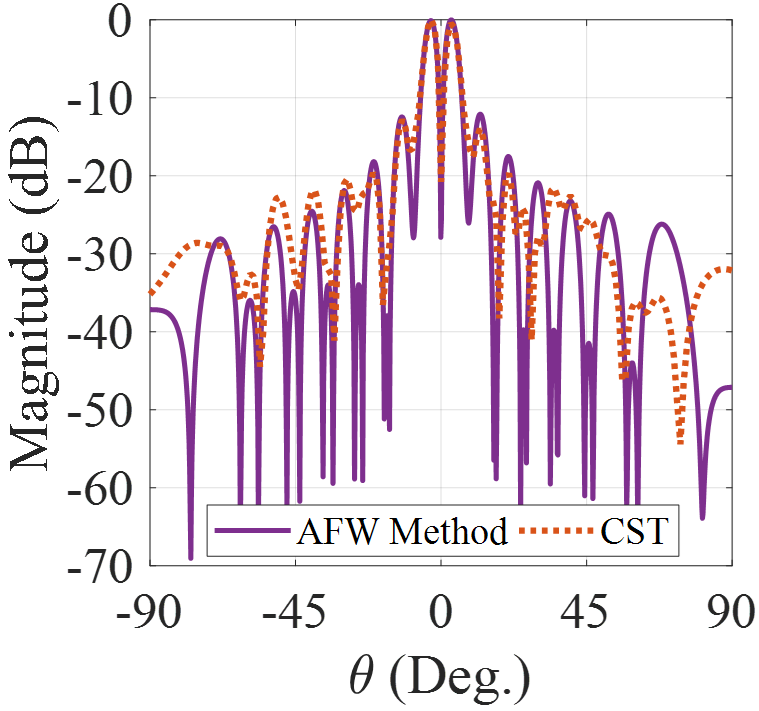}
		\caption{}
		\label{fig:pattern_comp_aniso_phi90}
\end{subfigure}
\caption{(a) Anisotropic surface impedance distribution and realized model. (b) RHCP component of pattern at $\phi = 0^\circ$. (c) RHCP component of pattern at $\phi = 90^\circ$. }
\label{fig:imp_aniso}
\end{figure*}

\begin{figure*}
\begin{subfigure}[b]{0.3\textwidth}
		\includegraphics[width =\textwidth]{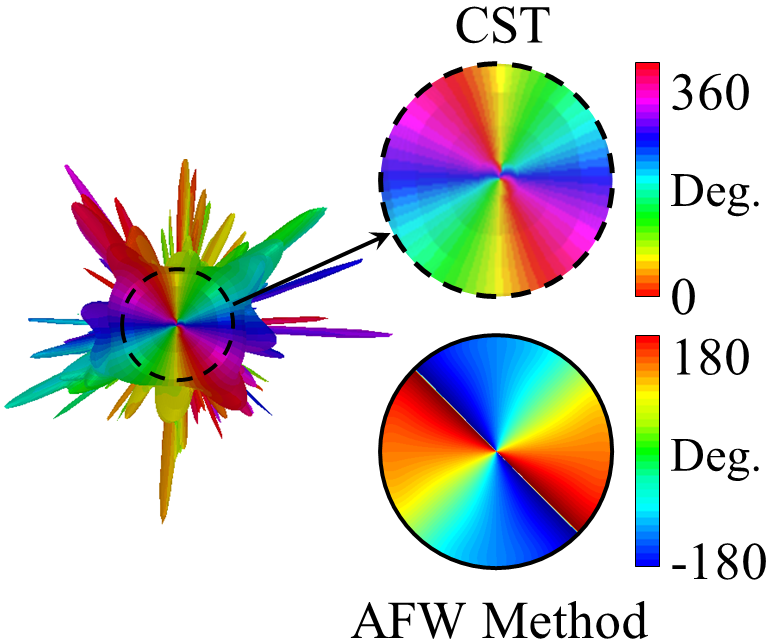}
		\caption{}
\end{subfigure}
\begin{subfigure}[b]{0.3\textwidth}
		\includegraphics[width =\textwidth]{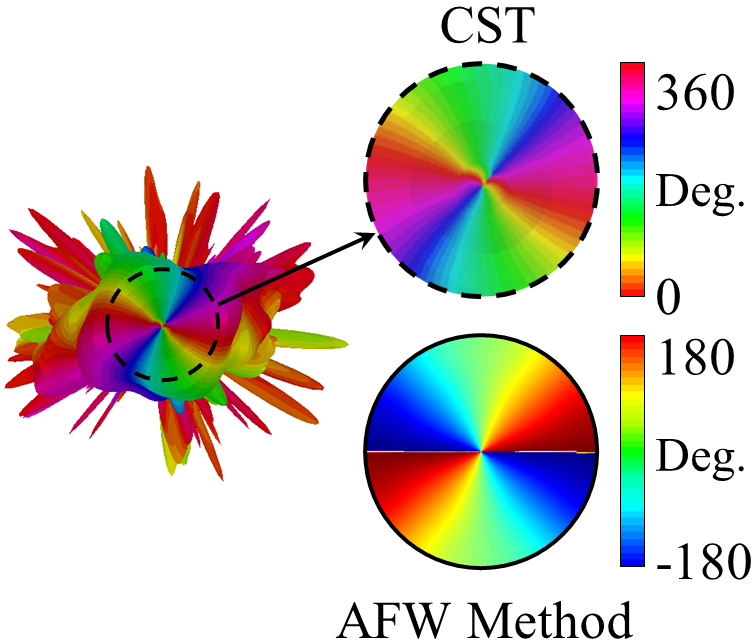}
		\caption{}
\end{subfigure}
\caption{Phase distributions of far-zone components. (a) $F_\phi$. (b) $F_\theta$, for anisotropic hologram in Fig.\ref{fig:imp_aniso}. }
\label{fig:F_aniso}
\end{figure*}

Unlike isotropic structures, in anisotropic holograms the components of field vector can be defined independently.  This enables us to control the polarization of radiation field. For example, to achieve circular polarization at ($\theta_0$, $\phi_0$) the horizontal and vertical components of far-zone field  must satisfy the following condition:
\begin{equation}
F_\phi(\theta_0, \phi_0) = e^{\pm j \frac{\pi}{2}} F_\theta(\theta_0, \phi_0)
\label{eq:F_phi_circ}
\end{equation}
where signs + and -  represent the left-hand and right-hand polarizations, respectively. Equation (\ref{eq:F_phi_circ}) imposes the following condition on aperture field components:
\begin{equation}
E_{ay}(\vec{\rho}) = E_{ax}(\vec{\rho})\frac{\cos \theta_0 \sin \phi_0 + e^{\pm j \pi/2} \cos \phi_0}{\cos \theta_0 \cos \phi_0 - e^{\pm j \pi/2} \sin \phi_0}
\end{equation}
If $E_{ax}$ is defined according to Eq. (\ref{eq:Eaxay}), the modulation indices must have the following relationship:
\begin{equation}
M_y(\vec{\rho})=M_x(\vec{\rho})\frac{\cos \theta_0 \sin \phi_0 + e^{\pm j \pi/2} \cos \phi_0}{\cos \theta_0 \cos \phi_0 - e^{\pm j \pi/2} \sin \phi_0}
\end{equation}
For example, to have radiation with right circular polarization (RHCP) in the broadside ($\theta_0=0^\circ$ and $\phi_0 = 0^\circ$), the following condition must be met:
\begin{equation}
M_y(\vec{\rho})=-jM_x(\vec{\rho})=-jM
\label{eq:My_circ}
\end{equation}
where M is assumed to be constant. 
Using (\ref{eq:FF_theta}) and (\ref{eq:FF_phi}) we can estimate the far-zone components as
\begin{equation}
\begin{split}
F_\theta(\theta, \phi) = \iint_{ap}\frac{ M J_{sw} e^{-j\phi}}{\sqrt{2\pi\rho \sqrt{(\beta_{um}^{(0)} + \delta \beta(\vec{\rho}))^2 + \delta \alpha^2 (\vec{\rho})}}}\times\\
e^{-\delta \alpha (\vec{\rho})} e^{jk\rho' \sin \theta \cos (\phi - \phi') } e^{-jl\phi'}\rho'd\rho'd\phi'
\end{split}
\label{eq:int_aniso1}
\end{equation}
\begin{equation}
\begin{split}
F_\phi(\theta, \phi) = \iint_{ap}\frac{-jMJ_{sw} e^{-j\phi} \cos \theta}{\sqrt{2\pi\rho \sqrt{(\beta_{um}^{(0)} + \delta \beta(\vec{\rho}))^2 + \delta \alpha^2 (\vec{\rho})}}}\times\\
e^{-\delta \alpha (\vec{\rho})} e^{jk\rho' \sin \theta \cos (\phi - \phi')} e^{-jl\phi'}\rho'd\rho'd\phi'
\end{split}
\label{eq:int_aniso2}
\end{equation}
By substituting $u=\phi - \phi'$ in (\ref{eq:int_aniso1}) and (\ref{eq:int_aniso2}) and simplifying them, it can be observed that the phase of $F_\theta$ and $F_\phi$ possess the angular dependence in the form of $(l+1)\phi$, which represents the topological charge of the beam.
Substituting (\ref{eq:My_circ}) in (\ref{eq:Phi_rho}) and (\ref{eq:Phi_phi}) results
\begin{equation}
\Phi_\rho(\vec{\rho}) = (l+1)\phi
\end{equation}
\begin{equation}
\Phi_\phi(\vec{\rho}) = (l+1)\phi + \frac{\pi}{2}
\end{equation}
Fig.\ref{fig:imp_aniso} shows the synthesized surface impedance distribution to generate OAM beam with right-hand polarization and topological charge of m = 2. The comparison between RHCP components of theoretical and simulation results are given in Figs \ref{fig:pattern_comp_aniso_phi0} and \ref{fig:pattern_comp_aniso_phi90}. The agreement between the results confirms the accuracy of theoretical model. 
Fig.\ref{fig:F_aniso} shows the phase of $F_\theta$ and $F_\phi$.
It can be observed that both $F_\theta$ and $F_\phi$ have the phase singularity in broadside with topological charge of m = 2. 
\begin{table*}[t!]
\caption{Computational performances of theoretical and full-wave simulations for the analysis of proposed holograms in Figs \ref{fig:imp_iso} and \ref{fig:imp_aniso}.}
\label{tab:table1}
\begin{tabular}{ccccc}
\hline\hline
Solver               & Analysis Method & Processor                                                                       & Allocated RAM & Time per Process                                                             \\ \hline
CST Microwave Studio & FIT             & \begin{tabular}[c]{@{}c@{}}Core i7 6850K\\ (3.6 GHz per processor)\end{tabular} & 64 GB         & \begin{tabular}[c]{@{}c@{}}Isotropic: $\approx$ 8 h\\ Anisotropic: $\approx$ 10 h\end{tabular}    \\
Our Code (in Matlab) & AFW             & \begin{tabular}[c]{@{}c@{}}Core i7 6500U\\ (2.5 GHz per processor)\end{tabular} & 12 GB         & \begin{tabular}[c]{@{}c@{}}Isotropic: $\approx$ 150 sec\\ Anisotropic:  $\approx$ 140 sec\end{tabular} \\ \hline
\end{tabular}
\end{table*}
\section{Computational performance of proposed method}
To clarify the advantage of theoretical model over the full-wave simulation, in this section, the computational performance of the two methods are compared.
Generally, leaky-wave holograms can be considered as traveling wave radiators. However, in order to have an appropriate radiation performance, they must be relatively large. The dimensions of these antennas to achieve high gain beams are about $10\lambda$ to $20\lambda$. On the other hand, the dimensions of unit cells to realize the hologram must be small enough ($<\lambda/5$) so that the surface can act as a homogenized impedance boundary condition. Therefore, the full-wave simulator applies dense meshing to be able to distinguish each inclusion, which greatly increases the simulation time.
 Table \ref{tab:table1} compares the CPU time for both full-wave and theoretical methods.
 The allocated RAM for the theoretical model is 12 GB. Also, the processor dedicated to it is Core i7 6500U (Ultra Low Voltage) with two real cores.
 The processor used in full-wave simulations is the Core i7 6850K model with 6 real cores and the dedicated RAM is 64 GB. 
 The results of Table \ref{tab:table1} show that the analysis time for the proposed method is significantly shorter than the full-wave simulations. Observe too that the proposed method can be easily developed on conventional computers, where the full-wave simulation is impractical.

\section{Conclusion}
In this work, an analytical method based on adiabatic floquet-wave expansion and aperture field estimation technique is proposed for the implementation of vortex beam radiators. To determine the aperture field, three points must be considered: 1- The amplitude of the field specifying the shape of the pattern; 2- The phase of the aperture field that determines the direction  and topological charge of OAM wave; 3- The relationship between the x and y components of the aperture field that controls the polarization of radiation beam.
According to the above considerations, two structures based on isotropic and anisotropic unit cells is proposed. For both structures, the aperture fields are estimated to achieve the desired patterns, polarizations and topological charges and are synthesized by the relevant unit cells. It has been shown that the anisotropic unit cell has the distinct advantage of polarization control over that of the isotropic one.
The advantage of using the analytical method is its high speed and less resource requirements and consumption. Furthermore, it has very good accuracy, which makes it suitable for analyzing and designing holograms with very large dimensions.

\section*{Appendix A:}
This appendix describes the calculation procedure of $\delta \alpha$ in detail. In (\ref{eq:aalpha}) the tensor $\bar{\bar{\chi}}$ is obtained from the following equation \cite{minatti_2016}:
\begin{equation}
\begin{split}
\bar{\bar{\chi}}=\chi_{\rho\rho}\hat{\rho}\hat{\rho} + \chi_{\rho\phi}\hat{\rho}\hat{\phi} + \chi_{\phi\rho}\hat{\phi}\hat{\rho} + \hat{\chi}_{\phi\phi}\hat{\phi}\hat{\phi}\\
= -(\bar{\bar{X}}^{(0)}+ j\bar{\bar{Z}}_{GF}^{(0)}) - j(\bar{\bar{z}}^{(+1)} + \bar{\bar{z}}^{(-1)})
\end{split}
\label{eq:chi}
\end{equation}
Note that $\bar{\bar{Z}}_{GF}^{(0)}$ is the tensorial Green's function of grounded dielectric slab evaluated at $\vec{k}=\beta_{um}^{(0)}\hat{\rho}$ and can be obtained as follows:
\begin{equation}
\bar{\bar{Z}}_{GF}^{(0)} = -j[\bar{\bar{X}}_0^{-1}(\beta_{um}^{(0)})+ \bar{\bar{X}}_G^{-1}(\beta_{um}^{(0)})]^{-1}
\label{eq:ZZ_GF}
\end{equation}
where
\begin{equation}
\bar{\bar{X}}_0(\beta_{um}^{(0)}) = -\frac{\eta_0\sqrt{(\beta_{um}^{(0)})^2 - k^2}}{k}\hat{\rho}\hat{\rho} + \frac{\eta_0 k}{\sqrt{(\beta_{um}^{(0)})^2 - k^2}}\hat{\phi}\hat{\phi}
\label{eq:XX_0}
\end{equation}
\begin{equation}
\begin{split}
\bar{\bar{X}}_G(\beta_{um}^{(0)}) = \\ \frac{\eta_0 \sqrt{\epsilon_r k^2 - (\beta_{um}^{(0)})^2}}{k \epsilon_r} 
\tan (h\sqrt{\epsilon_r k^2 - (\beta_{(um)}^{(0)})^2})\hat{\rho}\hat{\rho} \\
 + \frac{\eta_0 k}{\sqrt{\epsilon_r k^2 - (\beta_{(um)^{(0)}})^2}}\tan (h\sqrt{\epsilon_r k^2 - (\beta_{(um)}^{(0)})^2})\hat{\phi}\hat{\phi}
\end{split}
\label{eq:XX_G}
\end{equation}
In (\ref{eq:chi}) the tensor $\bar{\bar{X}}^{(0)}$ denotes the average reactance in the case of $m_\rho = m_\phi = 0$ and is defined as:
\begin{equation}
\bar{\bar{X}}^{(0)}=X_\rho \hat{\rho}\hat{\rho} + X_\phi \hat{\phi}\hat{\phi}
\end{equation}
Also, $\bar{\bar{z}}^{(+1)}$ and $\bar{\bar{z}}^{(-1)}$ have the following forms:
\begin{equation}
\bar{\bar{z}}^{(+1)} = \bar{\bar{X}}^{(-1)}.[\bar{\bar{Z}}_{GF}^{(-1)} - j\bar{\bar{X}}^{(0)}]. \bar{\bar{X}}^{(+1)}
\label{eq:z_plus}
\end{equation}
\begin{equation}
\bar{\bar{z}}^{(-1)} = \bar{\bar{X}}^{(+1)}.[\bar{\bar{Z}}_{GF}^{(+1)} - j\bar{\bar{X}}^{(0)}]. \bar{\bar{X}}^{(-1)}
\label{eq:z_minus}
\end{equation}
where
\begin{equation}
\begin{split}
\bar{\bar{X}}^{(+1)} = \frac{1}{2}e^{-jKs(\vec{\rho})}[m_\rho(\vec{\rho})e^{-j\Phi_\rho(\vec{\rho})}(X_\rho \hat{\rho}\hat{\rho} - X_\phi \hat{\phi}\hat{\phi}) + \\
m_\phi(\vec{\rho}) e^{-j\Phi_\phi(\vec{\rho})}X_\rho (\hat{\rho}\hat{\phi} + \hat{\phi}\hat{\rho})]
\end{split}
\end{equation}
\begin{equation}
\begin{split}
\bar{\bar{X}}^{(-1)} = \frac{1}{2}e^{+jKs(\vec{\rho})}[m_\rho(\vec{\rho})e^{+j\Phi_\rho(\vec{\rho})}(X_\rho \hat{\rho}\hat{\rho} - X_\phi \hat{\phi}\hat{\phi}) + \\
m_\phi(\vec{\rho}) e^{+j\Phi_\phi(\vec{\rho})}X_{\rho} (\hat{\rho}\hat{\phi} + \hat{\phi}\hat{\rho})]
\end{split}
\end{equation}
Note that, In (\ref{eq:z_plus}) and (\ref{eq:z_minus}), the tensors $\bar{\bar{Z}}_{GF}^{(+1)}$ and $\bar{\bar{Z}}_{GF}^{(-1)}$ are the Green's Functions of the dielectric slab at $\vec{k} = \beta^{(+1)}\hat{\rho}$ and $\vec{k} = \beta^{(-1)}\hat{\rho}$, respectively. Using (\ref{eq:ZZ_GF})-(\ref{eq:XX_G}) and substituting $\beta_{um}^{(0)}$ with $\beta^{(+1)}$ and $\beta^{(-1)}$ we can determine the tensors  $\bar{\bar{Z}}_{GF}^{(+1)}$ and $\bar{\bar{Z}}_{GF}^{(-1)}$.

 \FloatBarrier
\bibliographystyle{unsrt} 

\begin{thebibliography}{00}

\bibitem{thide2007}
B. Thidé, H. Then, J. Sjöholm, K. Palmer, J. Bergman, T. D. Carozzi, Ya. N. Istomin, N. H. Ibragimov, and R. Khamitova, Utilization of photon orbital angular momentum in the  low-frequency radio domain, Phys. Rev. Lett. 99, 087701 (2007).

\bibitem{chen2020}
R. Chen, H. Zhou, M. Moretti, X. Wang, and J. Li, Orbital angular momentum waves: generation, detection, and emerging applications, IEEE Commun. Surv. Tutor, 22, 840  (2020).

\bibitem{gao2014}
X. Gao, S. Huang, Y. Wei, W. Zhai, W. Xu, S. Yin, J. Zhou,
and W. Gu, An orbital angular momentum radio communication system optimized by intensity controlled masks effectively: Theoretical design and experimental verification, Appl. Phys. Lett. 105, 241109 (2014).


\bibitem{hui2015}
X. Hui, S. Zheng, Y. Hu, C. Xu, X. Jin, H. Chi, and X.
Zhang, Ultralow reflectivity spiral phase plate for generation of millimeter-wave OAM beam, IEEE Antennas
Wirel. Propag. Lett. 14, 966 (2015).

\bibitem{yi2019}
J. Yi, X. Cao, R. Feng, B. Ratni, Z. Jiang, D. Zhu, L. Zhu, A. De Lustrac, D. H. Werner, and S. N. Burokur, All-dielectric transformed material for microwave broadband orbital angular momentum vortex beam, Phys. Rev. Appl. 12, 024064 (2019).

\bibitem{yi2_2019}
 J. Yi, M. Guo, R. Feng, B. Ratni, L. Zhu, D. H. Werner, and S. N. Burokur,  Design  and  validation  of  an  all-dielectric  metamaterial medium for collimating orbital-angular-momentum vortex waves at microwave  frequencies, Phys.  Rev.  Appl. 12, 34060 (2019).

\bibitem{jiang2018}
Z. H. Jiang, L. Kang, W. Hong, and D. H. Werner, Highly efficient broadband multiplexed millimeter-wave vortices from metasurface-enabled transmit-arrays of subwavelength thickness, Phys. Rev. Appl. 9, 064009 (2018).

\bibitem{karimipour2019}
M. Karimipour, N. Komjani, and I. Aryanian, Holographic-inspired multiple circularly polarized vortex-beam generation with flexible topological charges and beam directions, Phys. Rev. Appl. 11, 054027 (2019).

\bibitem{yu2011}
N. Yu, P. Genevet, M. A. Kats, F. Aieta, J. P. Tetienne, F. Capasso, and Z. Gaburro, Light propagation with phase discontinuities: Generalized laws of reflection and refraction, Science 334, 333 (2011).

\bibitem{kildishev2013}
A. V. Kildishev, A.  Boltasseva, and V. M. Shalaev, Planar photonics with metasurfaces, Science 339, 1232009 (2013).

\bibitem{yu2014}
N. Yu, and F. Capasso, Flat optics with designer metasurfaces, Nat. Mater. 13, 139 (2014). 

\bibitem{checcacci1970}
P. Checcacci, V. Russo, and A. Scheggi, Holographic antennas, IEEE Trans. Antennas Propag. 18, 811 (1970).

\bibitem{fong2010}
B. H. Fong, J. S. Colburn, J. J. Ottusch, J. L. Visher and D. F. Sievenpiper, Scalar and tensor holographic artificial impedance surfaces, IEEE Trans. Antennas Propag. 58, 3212  (2010).

\bibitem{oraizi2020}
H. Oraizi, H. Emamian, Generation of orbital angular momentum modes via holographic leaky-wave metasurfaces, Sci. Rep. 10, 7358 (2020). 

\bibitem{amini_2020}
A. Amini, H. Oraizi, M. Hamedani, and A. Keivaan, Wide-band polarization control of leaky waves on anisotropic holograms, Phys. Rev. Appl. 13, 014038 (2020).

\bibitem{meng2019}
X. Meng, J. Wu, Z. Wu, L. Yang, L. Huang, X. Li, T. Qu, and Z. Wu, Generation of multiple beams carrying different orbital angular momentum modes based on anisotropic holographic metasurfaces in the radio-frequency domain, Appl. Phys. Lett. 114, 093504 (2019).


\bibitem{bodehou_OAM_2019}
M. Bodehou, C. Craeye, and I. Huynen, Near-field shaping by leaky-wave metasurfaces: OAM and bessel beams synthesis, in \textit{2019 International Conference on Electromagnetics in Advanced Applications (ICEAA)} (IEEE, Granada, Spain, 2019), p. 1210.

\bibitem{casaletti_2017}
M. Teniou, H. Roussel, N. Capet, G. Piau, and M. Casaletti, Implementation of radiating aperture field distribution using tensorial metasurfaces, IEEE Trans. Antennas Propag. 65, 5895 (2017).

\bibitem{ovejero2015}
D. González-Ovejero, and S. Maci, Gaussian ring basis functions for the analysis of modulated metasurface antennas,  IEEE Trans. Antennas Propag. 63, 3982 (2015).

\bibitem{bodehou2019}
M. Bodehou, D. González-Ovejero, C. Craeye, and I. Huynen, Method of moments simulation of modulated metasurface antennas with a set of orthogonal entire-domain basis functions, IEEE Trans. Antennas Propag. 67, 1119 (2019).

\bibitem{minatti_2015}
G. Minatti, M. Faenzi, E. Martini, F. Caminita, P. De Vita, D. Gonzalez-Ovejero, M. Sabbadini, and S. Maci, Modulated metasurface antennas for space: Synthesis, analysis and realizations, IEEE Trans. Antennas Propag. 63, 1288 (2015).

\bibitem{bodehou2020}
M. Bodehou, E. Martini, S. Maci, I. Huynen, and C. Craeye, Multibeam and beam scanning with modulated metasurfaces,  IEEE Trans. Antennas Propag. 68, 1273 (2020).

\bibitem{ovejero2017}
D. González-Ovejero, G. Minatti, G. Chattopadhyay, and S. Maci, Multibeam by metasurface antennas, IEEE Trans.  Antennas Propag. 65, 2923 (2017).

\bibitem{faenzi2019}
M. Faenzi, G. Minatti, D. González-Ovejero, F. Caminita, E. Martini, C. Della Giovampaola, and S. Maci, Metasurface antennas: New models, applications and realizations, Sci. Rep. 9, 10178 (2019).

\bibitem{minatti_2016} 
G. Minatti, F. Caminita, E. Martini, and S. Maci, Flat optics for leaky-waves on modulated metasurfaces: Adiabatic Floquet-wave analysis, IEEE Trans. Antennas Propag. 64, 3896 (2016).

\bibitem{minatti_2_2016}
G. Minatti, F. Caminita, E. Martini, M. Sabbadini, and S. Maci, Synthesis of modulated-metasurface antennas with amplitude, phase, and polarization control, IEEE Trans. Antennas Propag. 64, 3907 (2016).

\bibitem{moeini_scirep_2019}
M. M. Moeini, H. Oraizi, A. Amini, and V. Nayyeri, Wide-band beam-scanning by surface wave confinement on leaky wave holograms, Sci. Rep. 9, 13227 (2019).


\bibitem{tretyakov_2003}
S. Tretyakov, \textit{Analytical modeling in applied electromagnetics} ( Artech House, Inc., Norwood, MA, 2003).


\bibitem{oliner_1959}
A. Oliner, and A. Hessel, Guided waves on sinusoidally-modulated reactance surfaces, IRE Trans. Antennas Propag. 7, 201 (1959).

\bibitem{patel_2011}
A. M. Patel, and A. Grbic, A printed leaky-wave antenna based on a sinusoidally-modulated reactance surface, IEEE Trans. Antennas Propag. 59, 2087 (2011).

\bibitem{balanis_2016}
C. A. Balanis, \textit{Antenna Theory: Analysis and Design} (John Wiley \& Sons, New York, 2016), 4th ed.

\bibitem{cst}
CST Studio Suite, Computer Simulation Technolog ag, https://www.cst.com/.

\bibitem{werner_2014}
D. H. Werner, and D. H. Kwon, \textit{Transformation Electromagnetics and Metamaterials: Fundamental Principles and Applications} (Springer-Verlag, London, 2014).

\bibitem{vaity_2015}
P. Vaity and L. Rusch, Perfect vortex beam: Fourier trans-
formation of a bessel beam, Opt. Lett. 40, 597 (2015).

\bibitem{minatti_2011}
G. Minatti, F. Caminita, M. Casaletti, and S. Maci, Spiral leaky-wave antennas based on modulated surface impedance, IEEE Trans. Antennas Propag. 59, 4436 (2011).



\end{thebibliography}

\end{document}